# 'Animal spirits' and expectations in nonlinear U.S. recession forecasting

Elliott Middleton
*1604 Birchbrook Drive, Flower Mound, TX 75028*
*E-mail:* elliott.middleton@aya.yale.edu .

**A two-variable model is developed to forecast the probability of recession in the U.S. economy. One variable is an adaptation-level theoretic "animal spirits" or confidence-level indicator based on the unemployment rate [1]. The second is the difference between 10-year and 1-year maturity Treasury interest rates, a proxy for cyclical expectations [2,3]. Under conditions of partial ignorance or bounded rationality, agents may seek confirmation of expectations using adaptation-level-based heuristics in making decisions [4]. In the U.S., confidence tends to collapse when the unemployment rate ceases to be better than its recent average in memory. Like many others [5], the model uses data a year or more old to explain movements of a dichotomous dependent variable for recession as defined by the U.S. National Bureau of Economic Standards (NBER) to show that the probability of recession in the US economy has been largely predetermined a year in advance. The innovation of the present effort is the introduction of a confidence variable, which appears to increase the qualitative accuracy and structural stability of the model in validation testing compared to others. The result suggests that adaptation level effects are pervasive in economic dynamics, as they are in physiology, psychology and psychophysics [6], and that a primary cause of the ending of the expansion phase of the business cycle may be the impossibility of ever-improving values of some bounded economic variables, and a consequent loss of confidence.**

The selection of variables representing confidence and expectations was motivated by two considerations. First, prior work had established the importance of the relationship between the current US unemployment rate and agents' adaptation level to the unemployment rate in the determination of survey confidence measures as well as measures of output. Second, the slope of the Treasury yield curve (the graph of yield to maturity against term to maturity) has been widely regarded by practitioners and academics for some time as a reliable guide to the proximate onset of recession; see [5] for a comprehensive review. Both measures are based on publicly available data likely to be available in the future.

John Maynard Keynes popularized the use of the term "animal spirits" in economics [7]. Here the term is taken to represent general confidence levels. Max Wilhelm Wundt discovered the importance of adaptation level [8]. Helson [6] and others since have pointed out that adaptation level influences virtually all perception and cognition. To agents in the large, flexible labor markets of the United States, the published unemployment rate carries huge psychological significance. Other things equal, it is probably the best single indicator of the probability of job loss for a working agent. Let

$$A_t = -\frac{U_t - \overline{U}_t}{\overline{U}_t} \tag{1}$$

where $A_t$ is the "animal spirits" indicator; $U_t$ is the unemployment rate; $\overline{U}_t$ is an exponential moving average representing the adaptation level [1] given by

$$\overline{U}_t = \frac{1}{\sum_{\ell=0}^{47} e^{-0.03\ell}} \sum_{\ell=0}^{47} e^{-0.03\ell} U_{t-\ell} . \tag{2}$$

where $\ell$ is the number of periods lagged prior to the current one. Confidence is assumed to vary inversely with the proportionate divergence of the current unemployment rate from the adaptation level. The minus sign reflects the fact that confidence is promoted when the unemployment rate is lower than adaptation level and conversely. If the measure were to be based upon an output variable such as Industrial Production, to which agents were sensitive, there would be no minus sign; in fact, such a measure generates results qualitatively similar to, but not quite as robust, as those reported here. Figure 1 shows the graph of $A_t$, along with the slope of the yield curve.

$A_t$ has distributional properties that differ from those of $U_t$. Figure 2 shows log-log plots of the density of observations (y) against the absolute value of the first differences (x), that is, the absolute value of the current period value minus the previous period value, for both $U_t$ and $A_t$. This technique is used to examine $1/f^b$ distributional processes for universality, which is evident if a straight line adequately approximates the log-log graph [9]. Many loosely coupled natural and economic processes with a tendency toward aperiodic collapses or self-organized criticality are characterized in this way [10]. The distribution of changes in $U_t$ appears to fit a straight line, the distribution of changes in $A_t$ does not. There is an organized process of lower probability, larger than average movements occurring around business slumps. $A_t$ demonstrates a tendency to move most sharply when dropping below zero from above. It appears that as $U_t$ bottoms out, as it must, and as the adaptation level falls within a just noticeable difference [11] of the actual rate (and as $A_t$ approaches zero), agents' confidence dissipates. This emotional confirmation apparently validates the prior expectation of a slump. The expectation now becomes a self-fulfilling prophecy as production and consumption are curtailed due to the loss of confidence, further reducing confidence. Negative values of $A_t$ are strongly associated with recessions: the psychological dimension and the economic dimension are highly correlated. This result is also found with survey measures of confidence.

The University of Michigan index of consumer sentiment displays similar distributional characteristics to those of $A_t$. The present indicator was used because it produced slightly better results in forecasting, and is explicitly adaptation-level theoretic.

The slope of the yield curve, implemented here as the difference between the 10-year and 1-year Treasury rates, has proved to be a reliable cyclical indicator. This study used the 1-year Treasury rate instead of the commonly used 90-day rate to mitigate the impact Fed policy actions, which disproportionately affect the very short end of the yield curve.

The yield curve is normally upward sloping to compensate for greater price and time risk with a long bond and for future inflation [11]. The yield curve is said to invert when short-term rates rise above long-term rates. From 1955-present, there has been *almost* a one-to-one correspondence between inversions of the yield curve and a subsequent slump. Slumps began an average 13.2 months after inversions (maximum 20 months, minimum 8 months, standard deviation, 5.2 months) over 7 cycles. There was one exception. In 1966-1967 an inversion occurred prior to the mid-Sixties growth pause. Industrial production growth ceased but did not quite turn negative. A prior inversion of the yield curve has been a necessary condition for a recession to begin over the sample period. The yield curve inverted sharply in 1929, prior to the Crash and Great Depression.

Expectations are believed to be a major influence on the slope of the yield curve. Its intercept, the risk-free rate, is largely determined by inflation expectations. When agents expect the economy to slow, perhaps as the result of tight monetary policy, and expect long-term rates to decline, borrowers crowd into the short-term end of the curve, pushing short-term rates above long-term rates. Tight monetary policy aimed at curbing rising inflation toward the end of the cyclical expansion may also raise short-term rates. As demand for all borrowing slows with the slowing of business activity, and perhaps with monetary easing at the short end as inflation subsides, short-term and long-term interest rates float down, short-term rates more than long-term, until the curve reestablishes itself at a lower level overall. Businesses return to borrowing for longer terms, eager to lock in rates that are expected to rise in the next expansion phase of the business cycle.

My hypothesis was that confidence and expectations comprise deep dynamics of the business cycle that might be stable enough to be useful in forecasting. Previous work reviewed by Stock and Watson [5] used interest rate term and quality spreads, stock returns, dividend yields and exchange rates on data from various countries. Stock and Watson found that (1) multivariate models were better than univariate models, (2) out-of-sample validations were superior to in-sample statistics, (3) different variables were effective in different countries at different times, and (4) asset prices were more useful in predicting output growth than inflation. The yield curve has the widest following among regressors in recession forecasting models, so it was chosen to accompany the confidence variable. I note in passing that a yield quality spread (e.g., the difference between yields on a junk and an investment-grade bond) is a standard proxy for risk-aversion, which, while similar to a confidence variable, is not exactly the same thing.

Following standard procedure for developing a forecasting model, estimations were run over truncated portions of the data, generating in-sample results. Because the dependent variable was dichotomous (0 = no slump, 1 = slump) the maximum likelihood logistic estimation method was used. Logistic estimation is robust in the presence of non-normality of the regressors [13]. An algorithm searched the lag space from 12 months to 24 months for fit-maximizing lags of the two variables [14]. Multiple lags were not allowed. Using deeply lagged variables obviates the need to forecast any explanatory variables.

The estimation intervals all began in 1955:1, where coverage of all series began; data ended in 2001:4. Estimations were designed to end prior to major cyclical episodes of the postwar period. The ending points were 1968:12, 1978:12, and 1988:12. The search algorithm settled on lags of 21, 24, and 24 months for $A_t$ over the three periods, respectively, and on a consistent 12 month lag for the yield curve variable. Both variables' coefficients were consistently negative. In all cases $p < 0.0003$ for significance. Log likelihood < -40.0 in all cases. The addition of an $N(0,0.1)$ random variable to the percentage unemployment rate series to simulate undoing data revisions had no qualitative impact on the results.

Out-of-sample tests were conducted using forecasts generated by the estimated parameters of the models. In the case of logistic regression, the forecast is an estimated probability of the dichotomous variable occurring. Forecasts were generated for the 10 years following the end of each estimation interval. Figure 3 depicts the results from 1969:1-1985:12, Figure 4 from 1986:1-2002:4. Also shown is the annual growth rate of seasonally adjusted Industrial Production, defined as the logarithm of the current period value minus the logarithm of the value 12 months ago. Note that a unit probability estimate of recession is given in advance of some earlier recessions, a characteristic of some yield curve models also.

Out-of-sample results were collected for hits for the ten years following each estimation interval (without re-estimation); a hit occurred if estimated $p > 0.5$ during a slump month. For the out-of-sample period of 30 years (1969:1-1998:12) the hit rate was 67.7 percent. When slump is defined as the annual growth rate of Industrial Production being less than zero the hit rate is 72.4 percent. Over the 30-year model validation period including 5 slumps the model advanced a signal ($p > 0.5$) approximately a year before every recession. In a practical, qualitative sense, the model sent no false signals. This represents a 100 percent hit rate, which compares favorably with the track record of macroeconomic forecasting, which has been weak at forecasting turning points in the economy [15]. Given the apparent usefulness of the model as estimated, no analysis of potential break points was undertaken; this remains as an area for future research.

A true out-of-sample test is at hand, as the model forecasts a slump in second half 2001 ($p \approx 0.6$). A survey of professional forecasters estimated a one-third chance of a slump in 2001 [16].

One apparent advantage of the present model compared to the Stock and Watson model, which has been the benchmark in this area for the past decade [17], is the present model's $p > 0.5$ forecast in the vicinity of the 1990-1991 recession, which the yield curve models forecast with $p < 0.5$, after robustly signaling the recessions of the 1970s and 1980s. The present model's signal prior to the relatively mild 1990-1991 recession was weak as well. It may be that the dichotomous dependent variable approach is sensitive to the severity of recessions, or that the severity of recessions is sensitive to their chances of occurrence! If so, then the forecasted 2001 recession in the U.S. would be somewhat more severe than the 1990-1991 episode, but less severe than the 1970s and 1980s slumps. This is an area for future research.

In summary, a two-variable nonlinear model based on confidence and expectations predicts U.S. cyclical turning points well over an extended period. The chosen behavioral representation of a complex economic system apparently opens a window on a strongly attracting dynamical process. The ongoing forecasting success of the model depends on the continuing stability of the underlying attractor. The model may also be made more robust by the inclusion of other variables used in the literature. In applications to other countries, if the hypothesis is correct that failures of confidence cause business slumps, the key will be in identifying the (appropriately-transformed) variables that drive confidence; [4] suggests using a survey method to see what variables agents pay attention to. Income or consumption measures are likely candidates.

Adaptation and learning apparently did not change the modeled attractor enough over almost half a century to vitiate its forecasting properties. Indeed, consistent patterns of adaptation by both public and private agents are defining characteristics of the attractor. Subsequent learning and behavioral change by either set of agents, structural change, or substantial biasing of the relevant data by some means, may be necessary to force migration to another behavioral attractor.

However, a business "cycle" and alternating levels of confidence seem inevitable, so long as confidence is largely determined in a adaptation-level theoretic way [18]. The alternation may itself satisfy a preference for novelty that leads to greater fitness through continual adaptive challenge [19], suggesting an optimal degree of cyclical fluctuation. [20] sketches a theory of adaptation-level theoretic economic dynamics, with distributional consequences.

**Acknowledgements.** The author thanks Professors Y.-C. Zhang and Gene Stanley for encouragement.

**Supplementary materials.** RATS program and data files are available at http://members.home.net/elliottmiddleton/buscycle.zip.

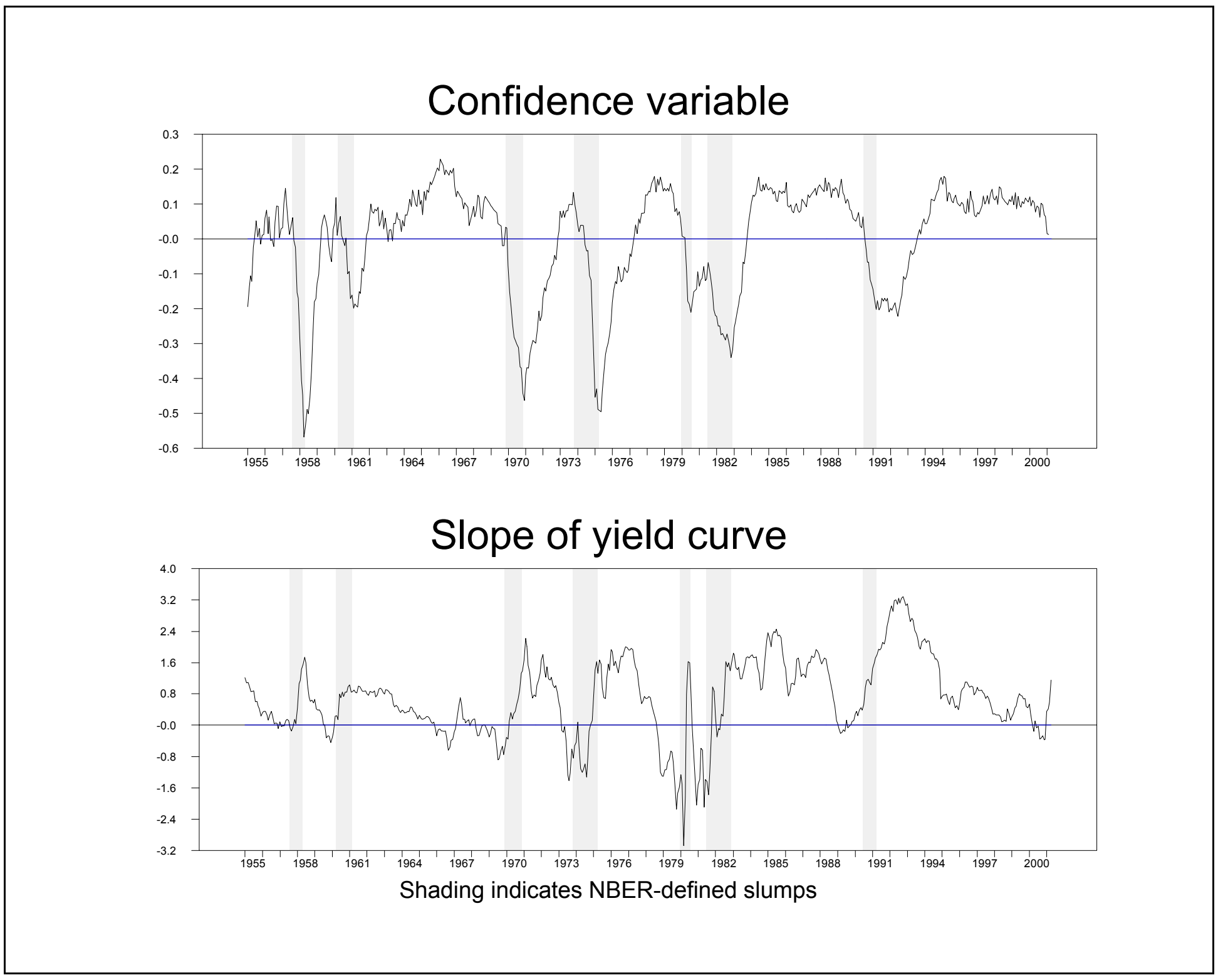

Figure 1. Model variables. Visually, the slope of the yield curve is a leading indicator, while *A* is a coincident indicator. Negative values of *A* are strongly associated with business slumps. Data through April 2001.

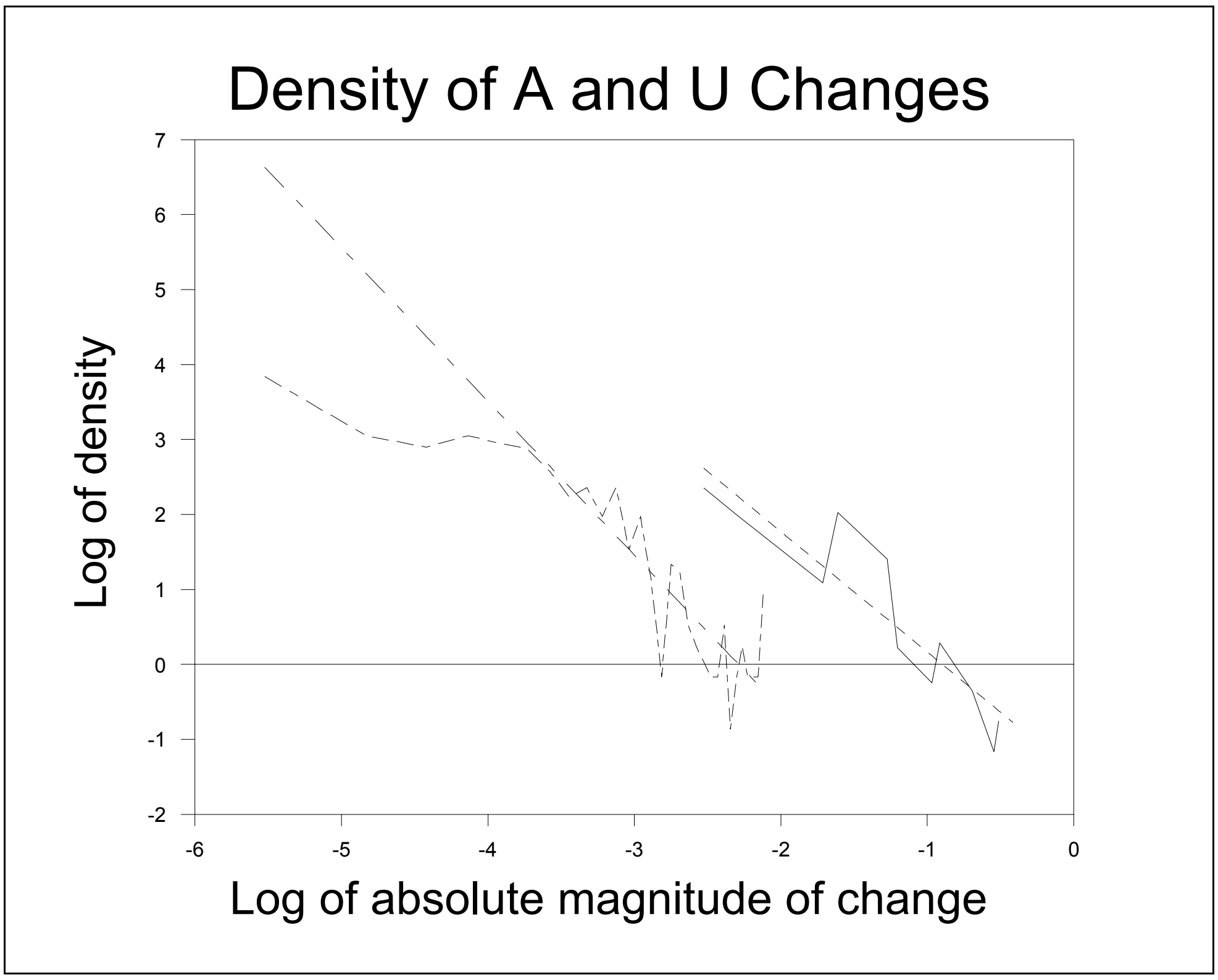

Figure 2. Scaling of changes in variables. Log-log plot for *A* (left) reveals that a different distributional process controls larger, lower probability movements than more common changes. Plot for *U* (right) reveals a single process.

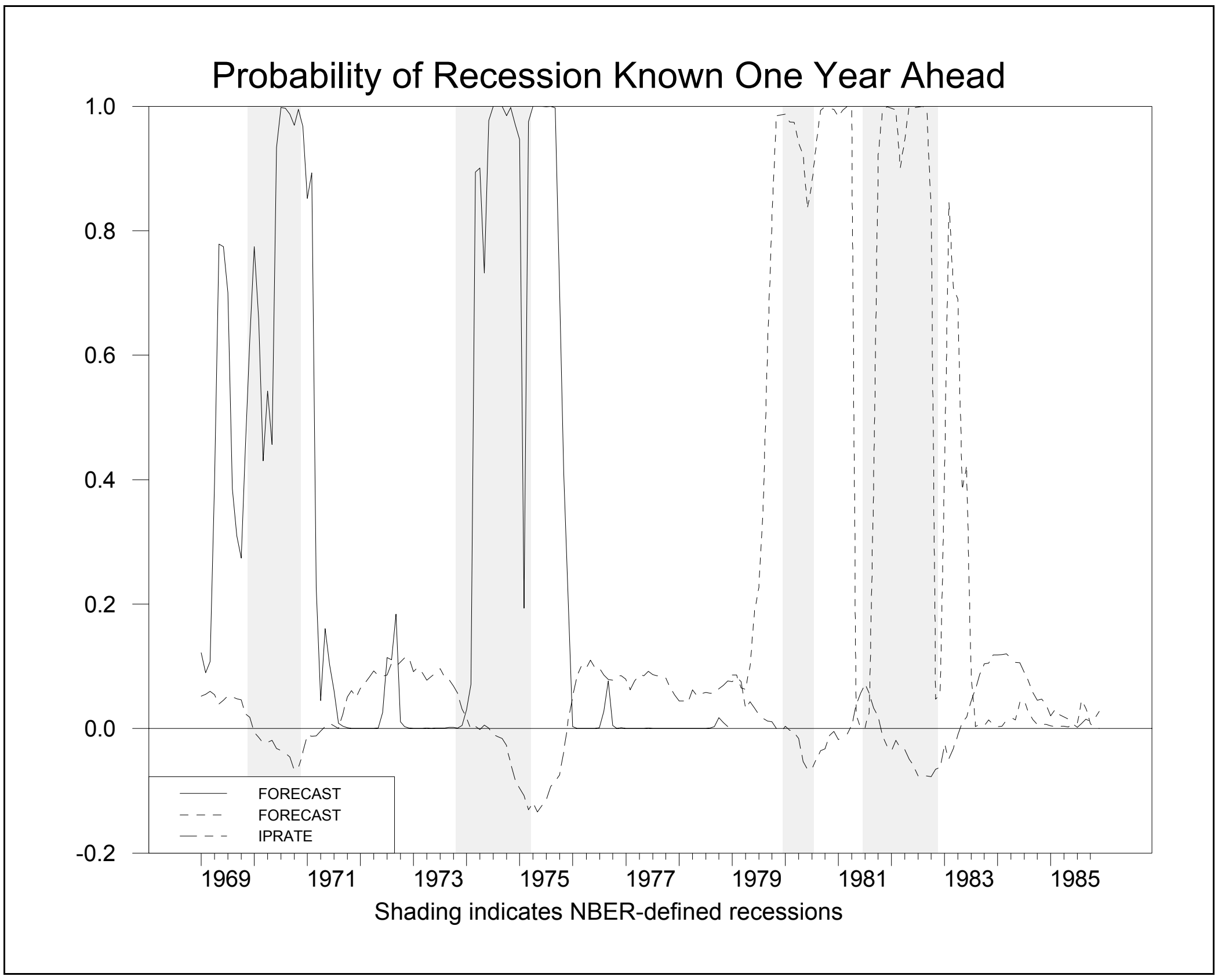

Figure 3. Validation testing. Forecasts are for the validation period 1969:1-1978:12 and first part of 1979:1-1988:12. Also shown is annual growth rate of Industrial Production. Data through April 2001.

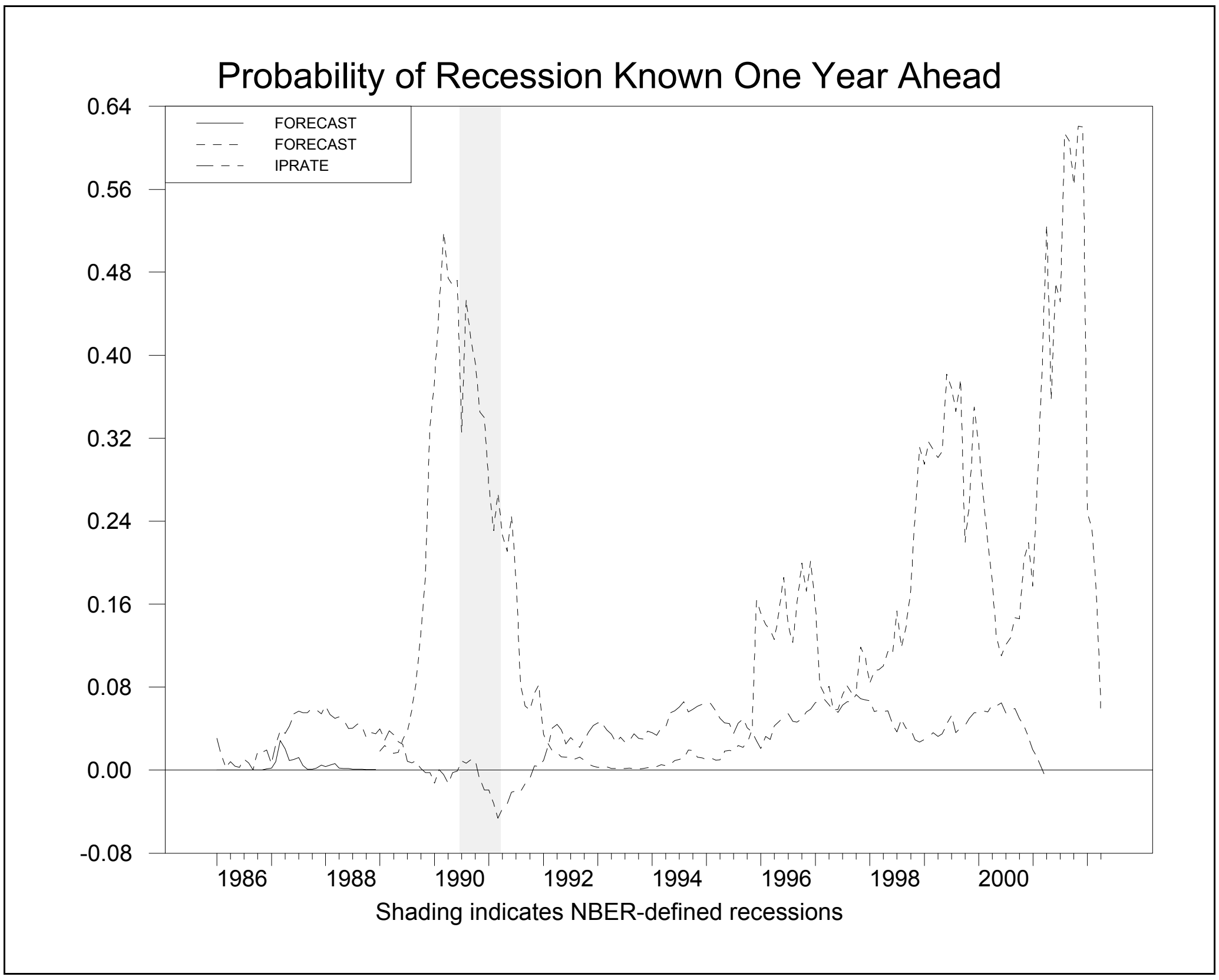

Figure 4. Validation testing. Second part of validation forecast 1979:1-1988:12, and continuation (without re-estimation) to generate current forecast for a slump in the second half of 2001 ($p \approx 0.6$), ending in first half 2002. Also shown is annual growth rate of Industrial Production. Data through April 2001.